\documentclass[twocolumn,preprintnumbers,showpacs,aps,prl,floats,amssymb]{revtex4}
\usepackage{graphicx}
\begin{document}
\author{Stefan Wessel$^{(1)}$}
\author{Matthias Troyer$^{(2)}$}
\affiliation{$^{(1)}$Institut f\"ur Theoretische Physik III, Universit\"at Stuttgart, 70550 Stuttgart, Germany}
\affiliation{$^{(2)}$Theoretische Physik, ETH Z\"urich, CH-8093 Z\"urich,  Switzerland}
\date{\today}
\title{Supersolid hardcore bosons on the triangular lattice}
\begin{abstract}
We determine the phase diagram of hardcore bosons on a triangular lattice with nearest neighbor repulsion, paying special attention to the 
stability of the supersolid phase. Similar to the same model on a square lattice we find that for densities $\rho<1/3$ or $\rho>2/3$ a supersolid phase is unstable and the transition between a commensurate solid and the superfluid is of first order. At intermediate fillings $1/3 < \rho < 2/3$ we find an extended supersolid phase even at half filling $\rho=1/2$.
\end{abstract}
\maketitle


Next to the widely observed superfluid and Bose-condensed phases with broken $U(1)$ symmetry  and ``crystalline'' density wave ordered phases with broken translational symmetry, the supersolid phase, breaking both the $U(1)$ symmetry and translational symmetry has been a widely discussed phase that is hard to find both in experiments and in theoretical models.
Experimentally, evidence for a possible supersolid phase in bulk $^4$He has recently been presented \cite{kc-nat}, but the question of whether a true supersolid has been observed is far from being settled \cite{leggett-2004,nikolay}, leaving the old question of supersolid behavior in translation invariant systems \cite{ss,anderson} unsettled for now. 

More precise statements for a supersolid phase can be made for bosons on regular {\em lattices}. It has been proposed that such bosonic lattice models can be realized by loading ultracold bosonic atoms into an optical lattice, where the required longer range interaction between the bosons could be induced by using the dipolar interaction in chromium condensates \cite{goral}, or an interaction mediated by fermionic atoms in a mixture of bosonic and fermionic atoms \cite{buechler}. With the recent realization of a Bose-Einstein condensate (BEC) in Chromium atoms \cite{pfau}, these experiments have now become feasible, raising the interest in phase diagrams of lattice boson model, and particularly in the stability of supersolids on lattices.
\begin{figure}
\includegraphics[width=8cm]{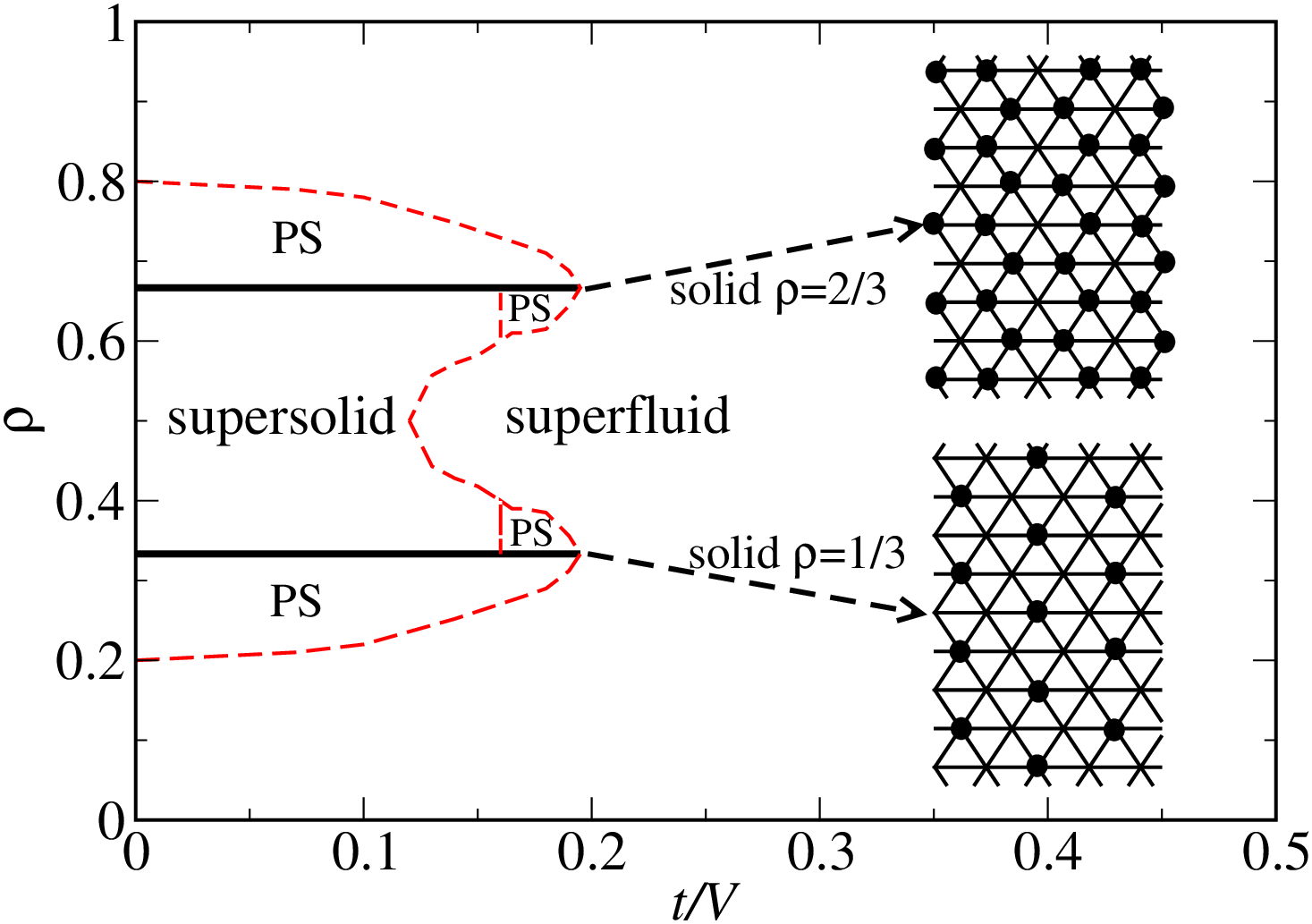}
\caption{Zero-temperature phase diagram of hardcore bosons on the triangular 
lattice in the canonical ensemble obtained from quantum Monte Carlo
simulations. The regions of phase separation are denoted by PS. 
The insets exhibit the density distribution inside the solid
phases for $\rho=1/3$ (lower panel), and  $\rho=2/3$ (upper panel).
}
\label{fig:canonic}
\end{figure}

The question if a supersolid phase is a stable thermodynamic phase for lattice boson models has been controversial for many years. Analytical calculations using mean-field and renormalization group methods \cite{mft1,balents,otterlo,batrouniold} have predicted supersolid phases for many models, including for the simplest model of hardcore bosons with nearest neighbor repulsion on a square lattice with Hamiltonian
\begin{equation}
H=-t \sum_{\langle i,j \rangle} \left( a^\dagger_i a_j + a^\dagger_j a_i \right) 
  +V \sum_{\langle i,j \rangle} n_i n_j 
  -\mu \sum_i n_i,
\label{eq:ham}
\end{equation}
where $a^\dagger_i$ ($a_i$) creates (destroys) a particle on site $i$, $t$ denotes
the nearest-neighbor hopping, $V$ a nearest-neighbor repulsion, and $\mu$ the chemical
potential. Subsequent numerical investigations using exact diagonalization and quantum Monte Carlo (QMC) algorithms \cite{batrouni,ps,schmid,guido,sengupta} have shown that for this model, the supersolid phase is unstable and phase separates into superfluid and solid domains at a first order (quantum) phase transition. Recently, this occurrence of a first order phase transition was explained by showing that a uniform supersolid phase in a hardcore boson model is unstable towards the introduction of domain walls, lowering the kinetic energy of the system by enhancing the mobility of the bosons on the domain wall \cite{sengupta}. 
In a related work it has been proposed that superfluid domain walls might be an explanation for the experimental observation of possible supersolidity in Helium \cite{nikolay,zhenya}.

To stabilize a supersolid on the square lattice, the kinetic energy of the bosons in the supersolid has to be enhanced either by sufficiently reducing the on-site interaction to be less than $4V$ \cite{sengupta},  by adding additional next-nearest-neighbor hopping terms \cite{guido}, or by forming striped solid phases with additional longer-ranged repulsions \cite{batrouni,schmid2}. 

In this Letter we will consider the interplay of supersolidity and {\it frustration}
by studying the hardcore boson model (\ref{eq:ham}) on a {\it triangular}
lattice. In the classical limit $t=0$ two solid phases exist at fillings
$\rho=1/3$ (and $\rho=2/3$), where one of three sites is filled (empty) in a
$\sqrt{3}\times\sqrt{3}$ ordering with wave vector ${\mathbf
  Q}=(4\pi/3,0)$ ~\cite{metcalf}, shown in the insets of Fig.~\ref{fig:canonic}. At half filling ($\rho=1/2$), where the square lattice shows a solid ordering with wave vector $(\pi,\pi)$, the solid order is frustrated on the triangular lattice, and the classical model has a hugely degenerate ground state with an extensive zero-temperature entropy~\cite{wannier}.

The question  arises whether this degeneracy of the classical system at half filling is lifted when quantum dynamics is added at a finite hopping parameter $t$, and which phase gets stabilized. Mean-field studies have predicted a supersolid phase \cite{murthy}. Given the questionable reliability of mean-field calculations in the case of the square lattice model a numerical check is needed. Indeed Green's function Monte Carlo (GFMC) simulations on small lattices \cite{boninsegni} have indicated the absence of a supersolid phase at half filling, but again on such small lattice numerical results can also be misleading as in the square lattice model \cite{batrouni}. 
\begin{figure}
\includegraphics[width=8cm]{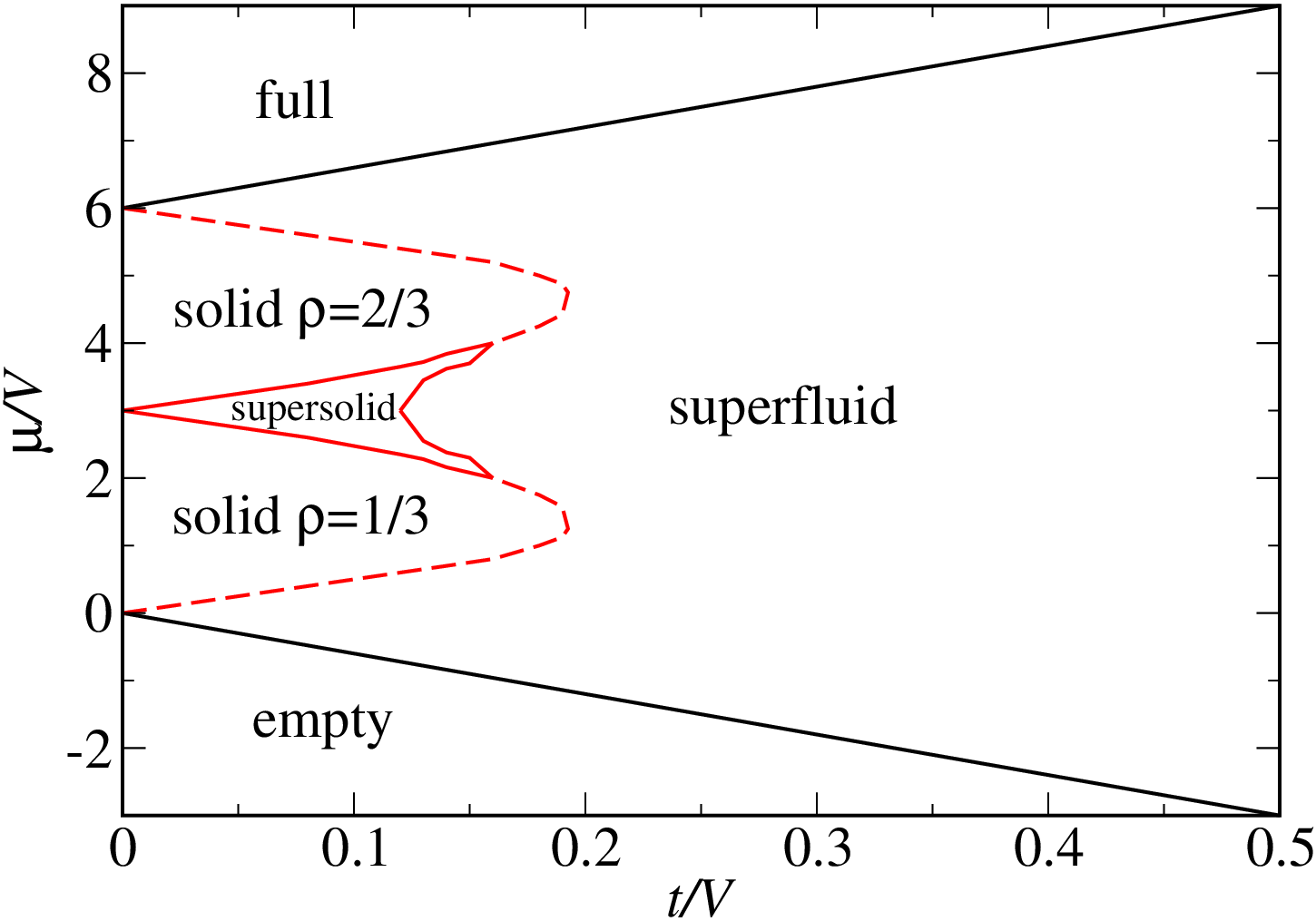}
\caption{Zero-temperature phase diagram of hardcore bosons on the triangular 
lattice in the grand canonical ensemble obtained from quantum Monte Carlo simulations. Second order 
phase transitions are denoted
by solid lines, whereas first-order transitions are denoted by dashed lines.
The system is half-filled for $\mu/V=3$.}
\label{fig:phasediag}
\end{figure}

We have thus performed a series of high-accuracy numerical QMC calculations on
large lattices using stochastic series expansions~\cite{sse} with  global directed-loop updates~\cite{directed} for the
hardcore boson model on the triangular lattice and show the phase diagram in
Fig.~\ref{fig:canonic} and Fig.~\ref{fig:phasediag} for the canonical and
grand-canonical ensemble, respectively. The main results are that for fillings
$\rho<1/3$ and $\rho>2/3$ a supersolid is unstable towards phase separation by
exactly the same domain-wall proliferation mechanism through which the square
lattice supersolid is unstable at all fillings $\rho\ne1/2$. In contrast, for
intermediate densities $1/3 < \rho < 2/3$ we find that the degeneracy of the
frustrated classical model is indeed lifted and a stable supersolid phase
emerges. The phase diagram in Fig.~\ref{fig:phasediag} is similar to the mean-field phase diagram~\cite{murthy}, albeit with a substantially reduced supersolid region. The supersolid is stable even at half filling, contradicting the small-lattice GFMC results of Ref. \onlinecite{boninsegni}.

We will now discuss the phase diagrams in more detail, starting with simple
limits. Considering the single boson (hole) problem, one can show that the lattice is empty for $\mu<\mu_{0}=-6t$ and completely filled for $\mu>\mu_{1}=6(t+V)$. For large values of  $t/V$, the bosons are superfluid, with a finite value of the superfluid density $\rho_S$, which we measure through the winding number fluctuations $W$ of the world lines~\cite{windingnumber} as
$\rho_S=\langle W^2 \rangle/({4 \beta t})$.
Two solid phases emerge upon lowering $t/V$ with rational fillings $1/3$ and $2/3$, respectively.
Both are characterized by a finite value of the density structure factor per site,
$S({\mathbf q})/N=\langle \rho_{{\mathbf q}} \rho^{\dagger}_{{\mathbf q}} \rangle$, where
$\rho_{{\mathbf q}}=(1/N)\sum_i n_i \exp(i {\mathbf q} {\mathbf r_i})$
at wave vectors $\pm{\mathbf Q}=\pm(4\pi/3,0)$, corresponding to the $\sqrt{3}\times\sqrt{3}$ ordering
wave vector. The maximum 
extent of the solid phases is reduced by quantum fluctuations from the mean-field value of $(t/V)_c=0.5$
down to $(t/V)_c=0.195\pm0.025$. 
\begin{figure}
\includegraphics[width=8cm]{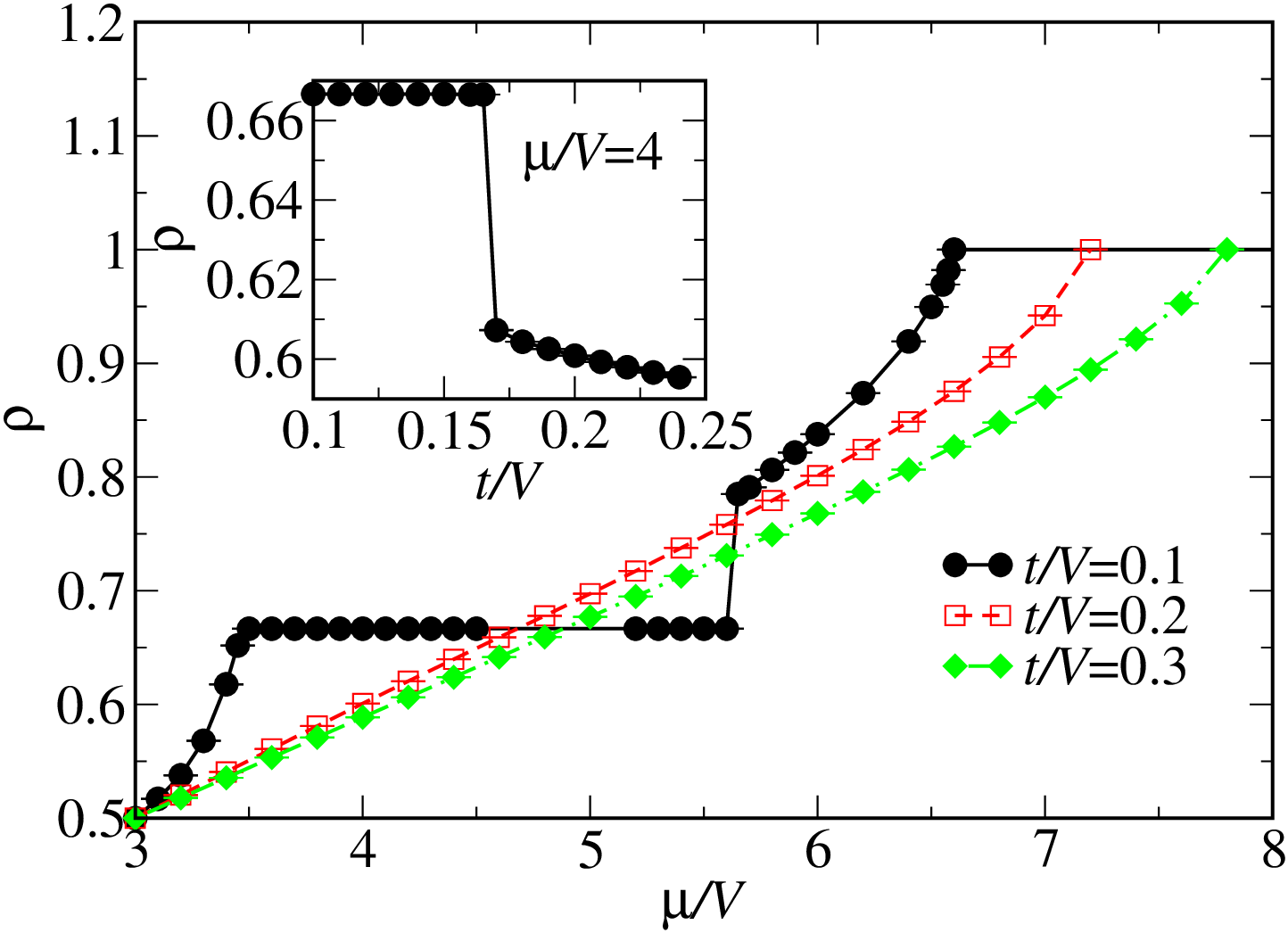}
\caption{Density of hardcore bosons on the triangular lattice as a function of $\mu$
along lines of constant values of $t/V$. The inset displays the jump in the
density as a function of $t$ for $\mu/V=4$ at $t/V\approx 0.165$. Only 
densities $\rho\ge 1/2$ are shown since the phase diagram is symmetric 
around half filling.} \label{fig:density}
\end{figure}
\begin{figure}
\includegraphics[width=8cm]{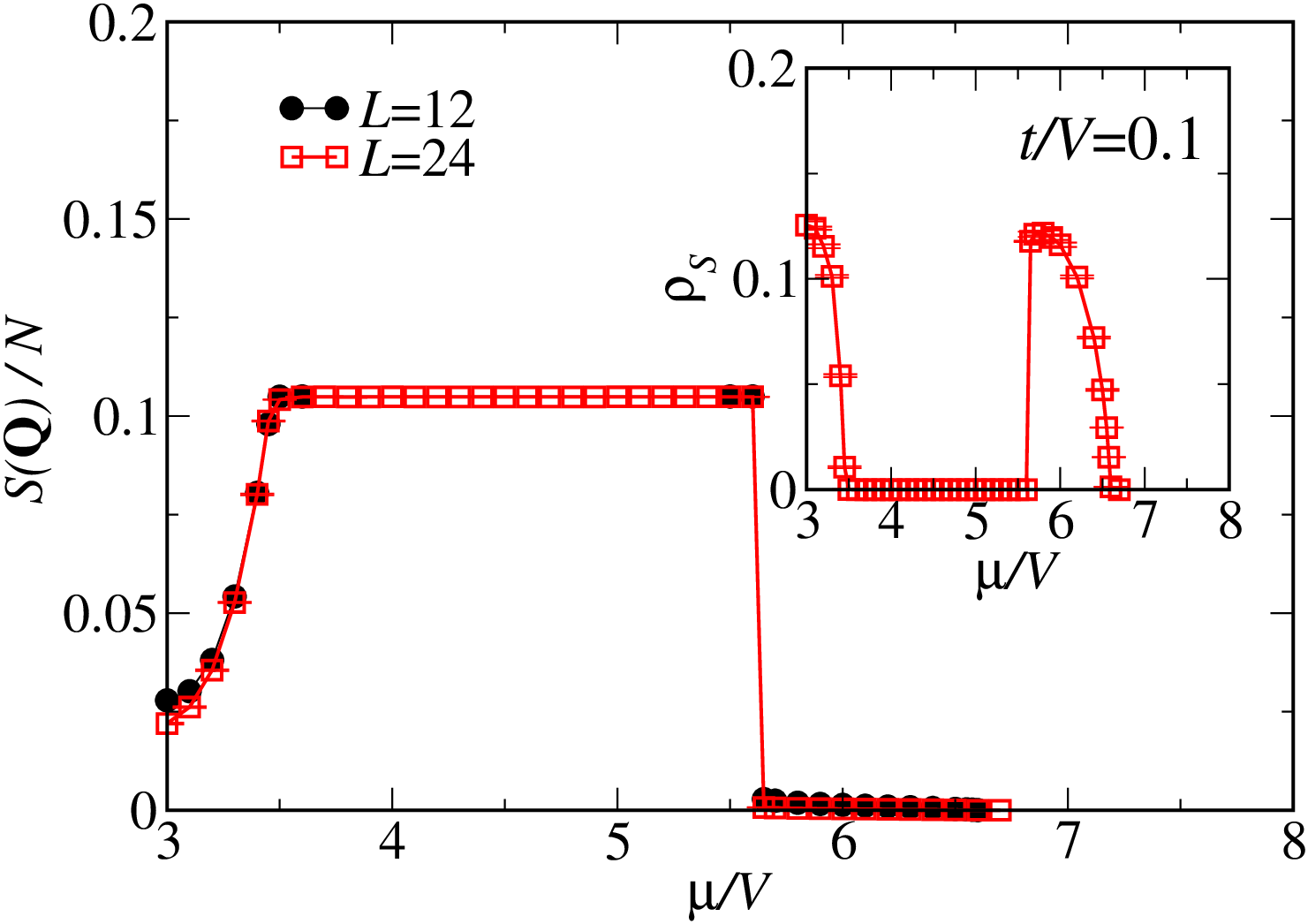}
\caption{
Static structure factor $S({\mathbf Q})$ for hardcore bosons on the triangular lattice as a function of $\mu$
along a line of constant $t/V=0.1$. The inset shows the behavior of the superfluid density $\rho_S$.
}
\label{fig:cut_t0.1}
\end{figure}

Since the phase diagram is symmetric when interchanging particles with holes ($\rho\rightarrow1-\rho$) 
we restrict our discussion from now on to $\rho\ge1/2$ and plot the density $\rho$ as a function of chemical 
potential $\mu$ for cuts at constant $t/V$ in Fig.~\ref{fig:density}.  For $t/V=0.1 $ we clearly observe a plateaux corresponding 
to the $\rho=2/3$ ($\rho=1/3$) 
phase with broken translational symmetry. The approach to this plateaux from $\rho<2/3$ ($\rho>1/3$) is continuous, indicating a second order phase transition, while for $\rho>2/3$ ($\rho<1/3$) we see a jump caused by a first order phase transition. Measuring the density structure factor $S({\mathbf q})$ and the superfluid density in Fig. \ref{fig:cut_t0.1} we identify this as a first order phase transition between the solid and superfluid phases. 
\begin{figure}
\includegraphics[width=8cm]{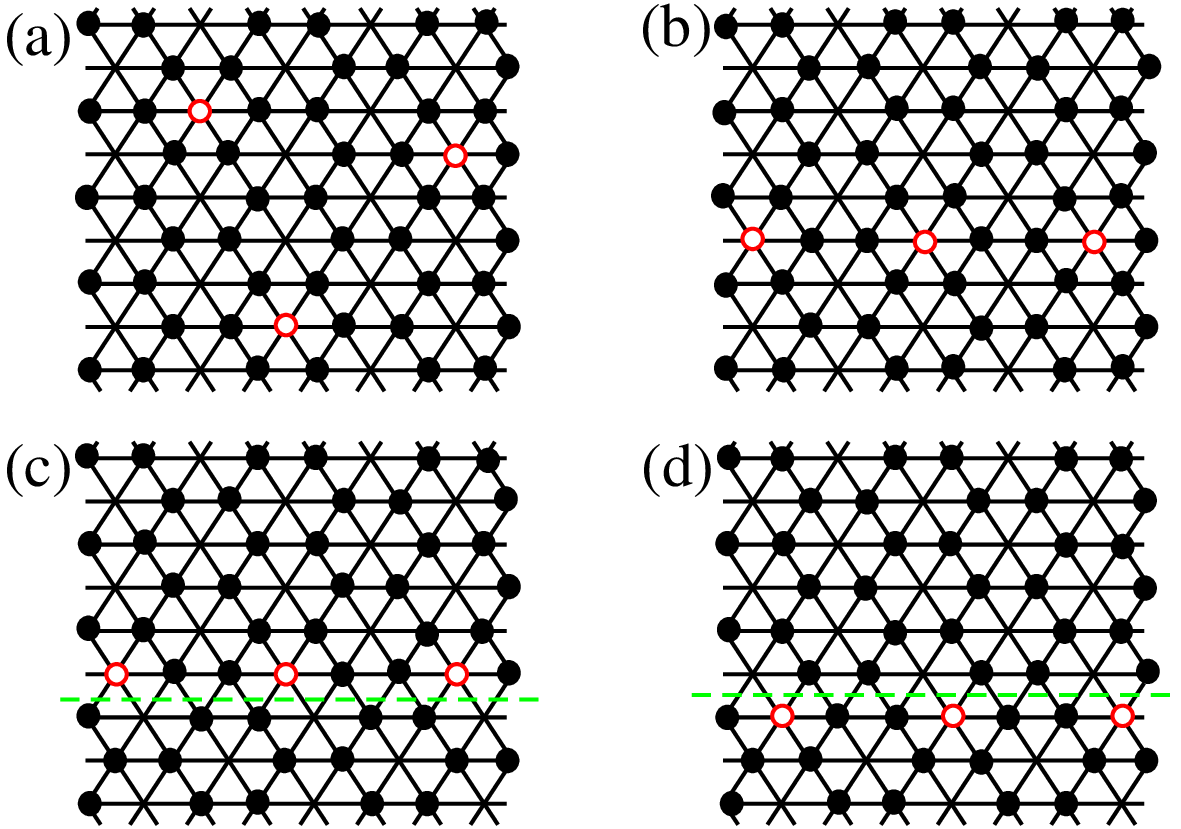}
\caption{
The $\rho=2/3$ solid doped with bosons. a) additional bosons (open circles) added on top of the solid. b) lining the bosons up costs no 
additional 
potential energy. c) shifting the lower half of the lattice introduces a domain wall (dashed line) at no cost, but now d) the additional 
particles can hop freely across the domain wall, gaining additional kinetic energy.}
\label{fig:instabel}
\end{figure}

The situation here is the same as in the square lattice model, where doping the solid leads to phase separation at a first order phase transition. The strict arguments for instability of a supersolid phase in the square lattice \cite{sengupta} can also be applied here: the uniform supersolid is unstable towards the introduction of domain walls as we illustrate in Fig. \ref{fig:instabel}. We start by adding $L/3$ additional bosons to the solid at density $\rho=2/3$ [Fig.  \ref{fig:instabel}a)], which corresponds to an infinitesimal density in the thermodynamic limit. These bosons can gain a kinetic energy of $-6t^2/V$ per boson by second order hopping processes. Placing these additional bosons along a line, as shown in Fig.  \ref{fig:instabel}b) costs no additional potential energy, and we can even shift one half of the lattice by one lattice spacing, introducing a domain wall as shown in Fig.  \ref{fig:instabel}c), again at no cost in potential energy. But now, the additional bosons can gain kinetic energy of $-t$ per boson by hopping freely across the domain wall, which lowers the energy of the domain wall state compared to the bulk supersolid, and hence the supersolid phase is unstable. 
\begin{figure}
\includegraphics[width=8cm]{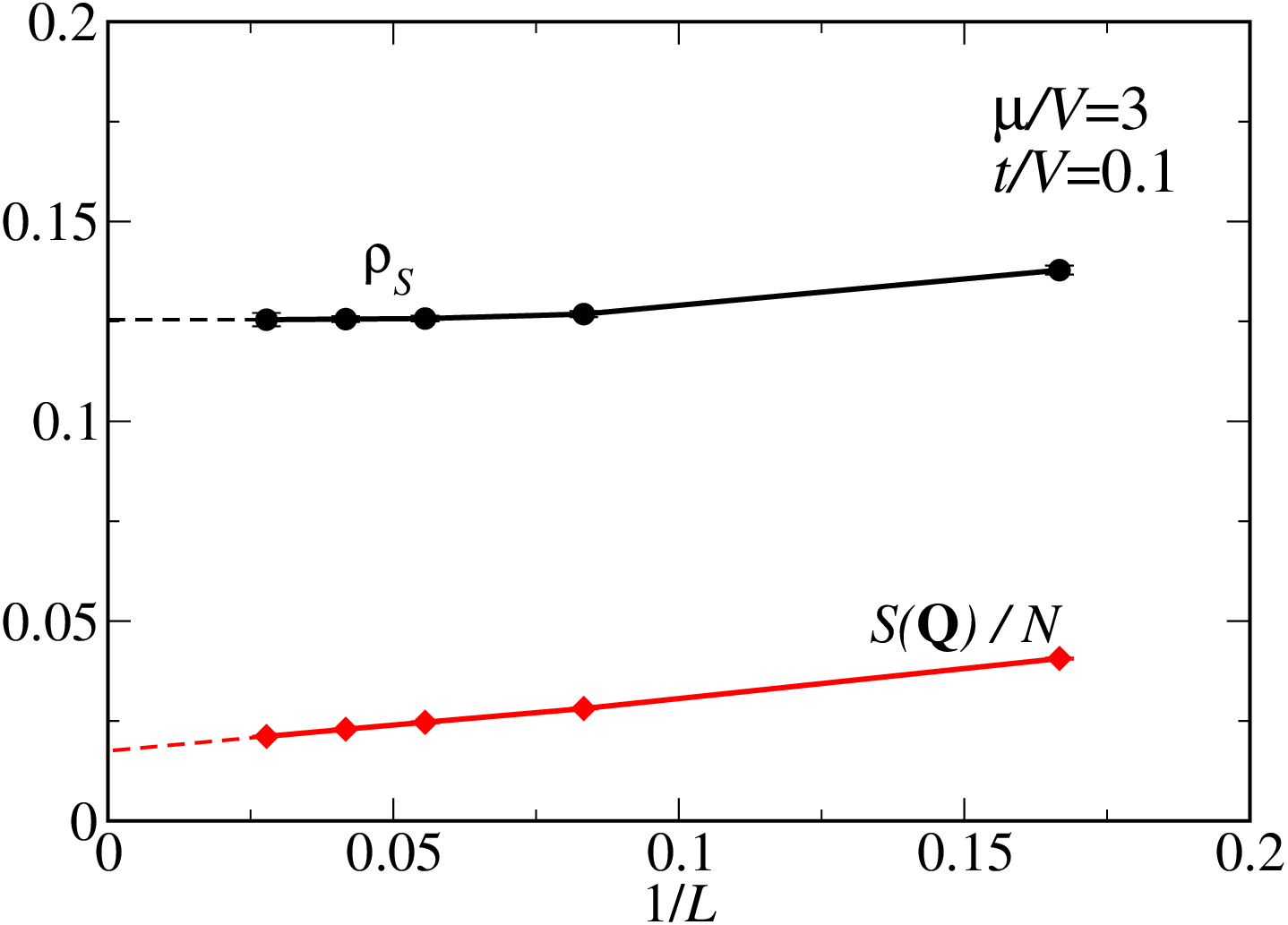}
\caption{
Finite size scaling behavior of the static structure factor $S({\mathbf Q})$ and the superfluid density $\rho_S$
for hardcore bosons on the triangular lattice at $t/V=0.1$ and half filling ($\rho=1/2$, $\mu/V=3$).
Dashed lines indicate extrapolations to the infinite lattice.
}
\label{fig:mu3_t0.1}
\end{figure}

\begin{figure}
\includegraphics[width=8cm]{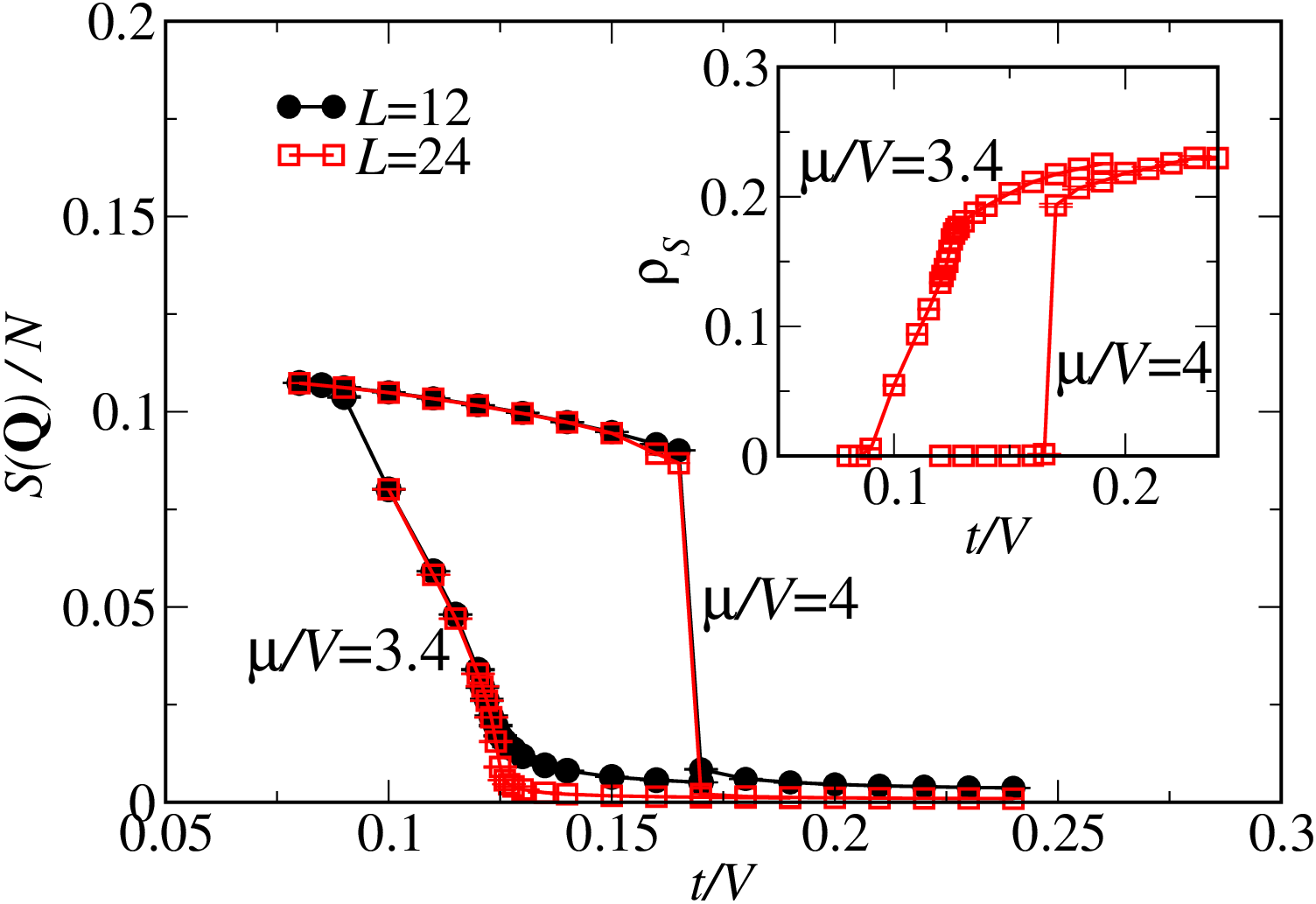}
\caption{
Static structure factor $S({\mathbf Q})$ for hardcore bosons on the triangular lattice as a function of $t$
along lines of constant $\mu/V=3.4$ and $\mu/V=4$. The inset shows the superfluid density $\rho_S$, exhibiting a kink
at $t/V\approx 0.125$ for $\mu/V=3.4$.
}
\label{fig:cut_mu4}
\end{figure}

A different situation exists for $\rho<2/3$, since there is no symmetry around $\rho=2/3$. 
Here, forming a domain wall would cost extra potential energy, and a supersolid phase can thus be stabilized. 
To demonstrate the existence of this supersolid even at half filling, we show the finite size scaling of 
$\rho_S$ and $S({\mathbf Q})$ in Fig. \ref{fig:mu3_t0.1}, both of which extrapolate to finite values.
\begin{figure}
\includegraphics[width=8cm]{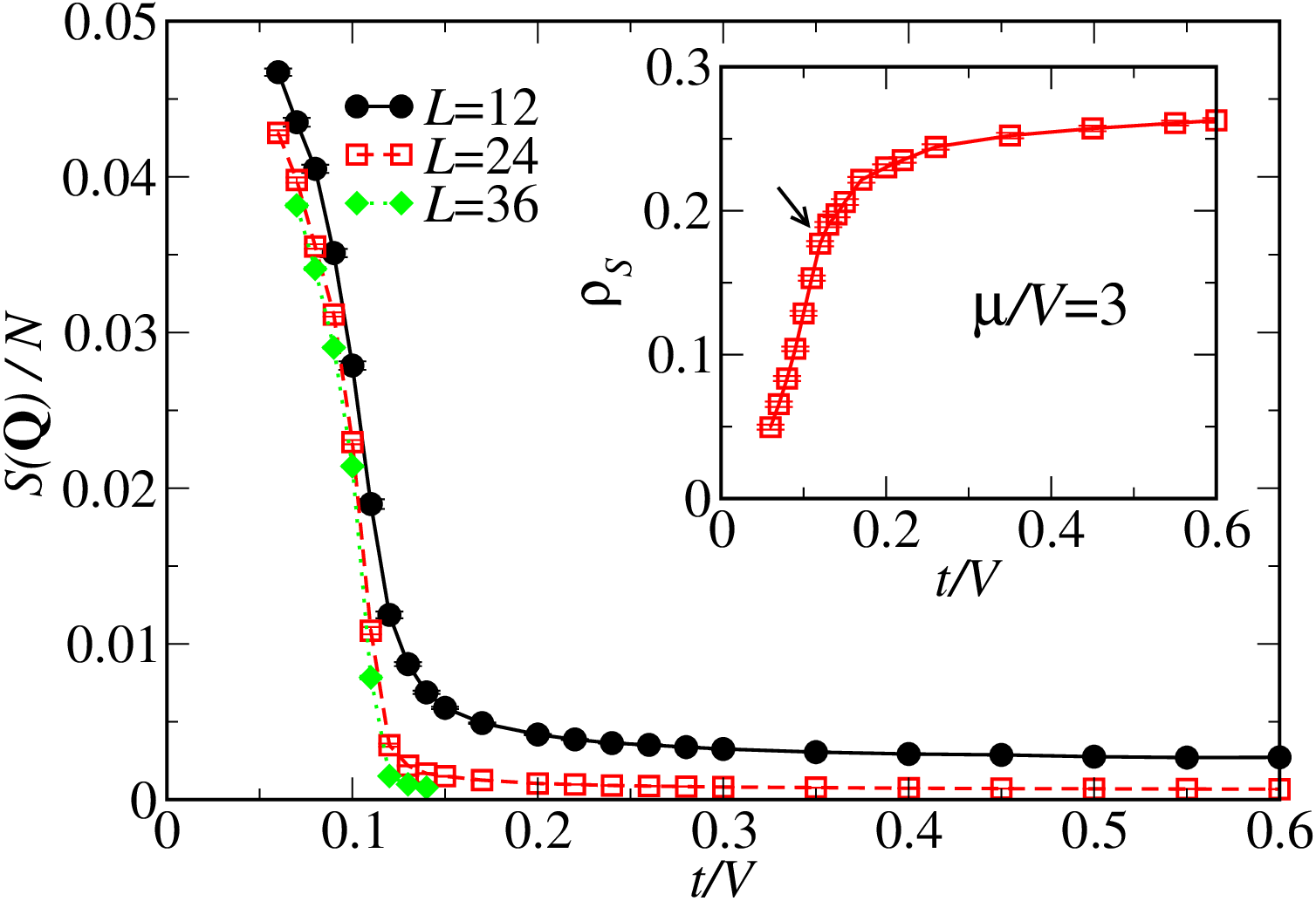}
\caption{
Static structure factor $S({\mathbf Q})$ for hardcore bosons on the triangular lattice as a function of $t$
at half filling ($\rho=1/2$ and $\mu/V=3$.) The inset shows the behavior of the superfluid density $\rho_S$ and the kink at $t/V\approx 
0.12$, indicated by an arrow.
}
\label{fig:cut_mu3}
\end{figure}
Intervening the solid phases at $1/3 < \rho < 2/3$ we hence find an extended supersolid phase, 
where both the superfluid density and the density structure factor take on finite values.
Fig.~\ref{fig:cut_mu3} shows $\rho_S$ and $S({\mathbf Q})$ as functions of $t/V$ at half 
filling, indicating a continuous quantum phase transition from the  supersolid to the superfluid at $t/V\approx0.115$.
We observe a kink in $\rho_S(t)$ near the transition point, marked by an arrow in Fig.~\ref{fig:cut_mu3}.
Away from half-filling, the extend of the supersolid phase slightly increases, as shown in Fig.~\ref{fig:phasediag}.
Moreover, the kink in $\rho_S(t)$ at the supersolid-superfluid transition becomes more pronounced, being clearly visible 
for $\mu/V=3.4$ in Fig.~\ref{fig:cut_mu4}.
Eventually, for $\mu/V>3.95$, the supersolid phase ceases to be stable, giving rise to a direct first order transition between the 
solid and the superfluid. This is reflected in discontinuities of both $\rho_S$ and $S({\mathbf Q})$ 
in Fig.~\ref{fig:cut_mu4}, as well as in the density $\rho$ [Fig.~\ref{fig:density}].
 
To summarize, we have demonstrated that, in contrast to square lattice hardcore boson models, an extended supersolid phase exists on the 
{\em 
triangular lattice} without the need for longer-range or softcore interactions, albeit in a smaller region than predicted by mean-field calculations \cite{murthy} and partially contradicting previous simulations on smaller lattices \cite{boninsegni}. This supersolid phase in the density regime $1/2 < \rho < 2/3$ emerges from the hugely degenerate disordered ground state of the frustrated classical model (in the $t=0$ limit) when the quantum mechanical hopping is turned on. This illustrates an intriguing mechanism by which a quantum system can avoid frustration: while $N/3$ of the bosons, on an $N$-site lattice form a non-frustrated solid at wave vector  $(4\pi/3,0)$ and break translational symmetry, the remaining $N(\rho-1/3)$ bosons delocalize and break the $U(1)$ gauge symmetry, forming a superfluid Bose-condensate on top of the solid with density $\rho=1/3$, thus realizing a supersolid phase. 

Since, in contrast to the square lattice, the triangular lattice model does not need additional longer-ranged repulsion or hopping terms, nor a reduction of the on-site interaction \cite{sengupta}, the triangular lattice might be preferred over the square lattice when looking for supersolid phases in ultra-cold atoms on optical lattices.

We thank A. Auerbach, R. Melko and A. Muramatsu for discussions and acknowledge support of the Swiss National Science Foundation and the hospitality of the Kavli Institute of Theoretical Physics in Santa Barbara, where parts of this work were carried out.

\end{document}